\documentclass{mn2e}

\usepackage{graphicx}
\usepackage{graphics}
\usepackage{hhline}
\usepackage{natbib2}
\usepackage{natbibmnfix}
\usepackage{amssymb}
\usepackage{amsmath}
\usepackage{amsopn}
\usepackage{lscape}
\bibpunct[,]{(}{)}{;}{a}{}{}

\newif\ifAMStwofonts

\newcommand{\hkpcDot}{\mbox{$h^{-1}$ kpc.} }

\newcommand{\hmsun}{\mbox{$h^{-1}$ $M_{\odot}$} }

\newcommand{\kms}{\mbox{km s$^{-1}$} }

\newcommand{\Mpc}{\mbox{Mpc} }
\newcommand{\MpcDot}{\mbox{Mpc.} }

\newcommand{\kpc}{\mbox{kpc} }
\newcommand{\kpcDot}{\mbox{kpc.} }

\newcommand{\msun}{\mbox{$M_{\odot}$} }
\newcommand{\msunDot}{\mbox{$M_{\odot}$.} }
\newcommand{\msunKC}{\mbox{$M_{\odot}$),} }

\newcommand{\Mrvir}{\mbox{$m_{\rmn{vir}}$} }
\newcommand{\MrvirCom}{\mbox{$m_{\rmn{vir}}$,} }

\newcommand{\Rrvir}{\mbox{$r_{\rmn{vir}}$} }

\newcommand{\RrvirDot}{\mbox{$r_{\rmn{vir}}$.} }

\newcommand{\sph} {{\sc sph}\ }

\def\la{\mathrel{\hbox{\rlap{\hbox{\lower4pt\hbox{$\sim$}}}\hbox{$<$}}}}
\def\ga{\mathrel{\hbox{\rlap{\hbox{\lower4pt\hbox{$\sim$}}}\hbox{$>$}}}}

\def\gsim{\ga}
\def\lsim{\la}

\newcommand{\bc}{\begin{center}}
\newcommand{\ec}{\end{center}}

\newcommand{\be}{\begin{equation}}
\newcommand{\ee}{\end{equation}}

\renewcommand{\vec}[1]{\overrightarrow{#1}}  
\newcommand{\vecm}[1]{$\overrightarrow{ #1}$}  
\newcommand{\vechm}[1]{\overrightarrow{\bf #1}} 
\newcommand{\vechmm}[1]{$\overrightarrow{\bf #1}$}

\newcommand{\tx}[1] {\rmn{#1}}

\newcommand{\xnu}{{x_{\nu}}}
\newcommand{\xp}{\vechm{X}}
\newcommand{\yp}{\vechm{Y}}
\newcommand{\zp}{\vechm{Z}}
\newcommand{\coordp}{\xp,\yp,\zp}
\newcommand{\coordpm}{$\coordp$}
\newcommand{\ptSZ}{p$^{2}$tSZ }
\newcommand{\pkSZ}{p$^{2}$kSZ }

\title{SZ polarisation as a probe of the intracluster medium}
\author[G. Lavaux, J. M. Diego, H. Mathis, J. Silk]{G. Lavaux$^{1}{}^{2}$\thanks{gxl@astro.ox.ac.uk}, J. M. Diego$^{1}$, H. Mathis$^{1}$, J. Silk$^{1}$
\\
$^{1}$University of Oxford, Astrophysics, Denys Wilkinson Building, Keble Road, Oxford OX1 3RH, UK \\
$^{2}$Ecole Normale Sup\'erieure de Cachan, 61 Avenue du Pr\'esident Wilson, 94230 Cachan, France}

\pagerange{\pageref{firstpage}--\pageref{lastpage}}
 
\begin{document}
\maketitle
\label{firstpage}



\begin{abstract}
We present high-resolution hydrodynamical simulations of the degree and direction of polarisation 
imprinted on the CMB by the Sunyaev-Zel'dovich effect in the the line of sight to massive galaxy clusters. 
We focus on two contributions to the electron rest-frame radiation quadrupole anisotropy in addition to the intrinsic 
CMB quadrupole that contributes most of the  induced CMB polarisation: 
the radiation quadrupole seen by electrons  due to their  
own velocity  in the plane normal to the line of sight, and the radiation quadrupole due 
to the thermal Sunyaev-Zel'dovich effect, which is generated by a previous scattering 
elsewhere in the cores of the local and nearby clusters. 
We show that inside the virial radius of a massive cluster, this latter effect, 
although being  second order in the optical depth, can reach the nK level of the former effect. These two effects 
can, respectively, constrain the projected tangential velocity and inner density profile of the gas, if they can be separated  
with multi-frequency observations.  As the information on the direction and magnitude of the 
tangential velocity of the ICM gas is combined with the temperature Sunyaev-Zel'dovich effect  
probing the projected radial velocity it will be possible to improve our understanding of the dynamics of the ICM. 
In particular, future polarisation observations may be able to trace out  the filamentary structure of the cooler and less dense 
ionized gas that has not yet been incorporated into massive clusters. We also discuss how the 
collapse of a cluster should produce a peculiar ring-like polarisation pattern 
which would open a new window to detect proto-clusters at high redshift. 
\end{abstract}




\section[]{Introduction}
\label{sec:Intro}

Recent measurements of the TE power spectrum of the intrinsic
polarisation of the CMB by \emph{WMAP} have put strong constraints on
the total optical depth $\tau$ to the last scattering surface and on
the possible nature of the sources responsible for
reionisation. These measurements have also provided impetus for the
further development of polarisation-sensitive receivers and for the
theoretical modelling of the foregrounds contaminating the CMB
polarisation.

On the other hand, new generation X-ray satellites such as
\emph{Chandra} or \emph{XMM-Newton} have produced a wealth of high
resolution maps of the intracluster medium (ICM) of nearby clusters. They support some aspects
of our previous picture of the hydrostatic state of the plasma, but
they invalidate others, and, most interestingly, bring out new
phenomena such as cold fronts.  However, it is well known that
detailed studies of the cluster ICM in the X-ray band constrain the
innermost parts, while the dynamics of the gas is also interesting at larger
radii where for example gas flowing in from filaments, pristine gas
accreting from the low-density regions of the Universe, or subclumps
of cold gas of merging units encounter the hot ICM.

The Sunyaev-Zel'dovich effect (hereafter, SZ) encapsulates the processes of Thomson and 
inverse Compton scattering of CMB photons by the free electrons of the ICM (see, e.g. \citealt{Sun80b,Carl02}).
In CMB temperature maps,  both the thermal SZ (hereafter tSZ) effect and the kinetic SZ (hereafter kSZ) effects contribute to the secondary anisotropies. 
With CMB temperature variations depending 
at fixed observing frequency only on the gas pressure integrated along the line-of-sight, the tSZ effect has long been recognized as a powerful probe for studying the morphology of the ICM 
out to a substantial fraction of the virial radius of the cluster.  The kSZ effect 
typically amounts to a few percent  of the thermal effect and maps the bulk  
velocity of the ICM  projected along the line of sight, to the extent that the ICM can be considered as a coherent bulk.
Multi-frequency observations are necessary to separate the kSZ effect  from the tSZ effect 
and extract information on the radial velocity of the gas.  Recently, 
using Eulerian hydrodynamical simulations, \citet{Nag03b} have studied   
 the bias due to the internal  motion of the gas in a massive, $2.4\times 10^{14}$\hmsun  
cluster on the  estimates of the cluster bulk velocity along the 
same lines of sight where  one measures the kSZ effect. While 
they focused on the ability to determine the global motions of massive haloes 
to later constrain cosmological parameters by means of the 
cluster pairwise peculiar velocities, it is also possible to constrain  
the motion of the ICM in the rest frame of the cluster. 
In fact, the kSZ effect and some of the SZ polarisation signals are complementary as they allow one to obtain information 
on, respectively, the projection along the line of sight of the density-weighted radial velocity of the gas,  
and the projection (amplitude and direction) along the line of sight of the 
density-weighted tangential velocity of the gas.

On CMB polarisation maps, the SZ-induced polarisation comes mainly
from four terms. The first, which we want to emphasize, comes from
single scattering of the quadrupole generated as radiation is
Doppler--shifted due to the tangential motion of the gas (hereafter
kpSZ), the second is due to single scattering of the CMB intrinsic
quadrupole (hereafter ipCMB), and the last two are second scatterings of
quadrupole anisotropies induced in the CMB by respectively previous t-
and kSZ effects induced by first scatterings (hereafter respectively
\ptSZ and \pkSZ), when the y parameter/optical depth seen by the last
scattering electron is not isotropic.  These last two effects are
potentially powerful probes of the central gas profiles in clusters.

For last scattering electrons inside and outside clusters, the \pkSZ effect
is negligible relative to the \ptSZ effect if clusters are sufficiently hot
and massive: these terms scale respectively as  $\eta \tau^2$ and $\beta \tau^2$
\citep[hereafter S99]{Saz99} where $\eta=kT/mc^2$ and $\beta$ the tangential velocity.
Indeed, massive clusters have typically  
$\eta\simeq 3\; 10^{-2}$ and $\beta\simeq 3\; 10^{-3}$.
While the ipCMB can reach levels of order 50 nK (S99), 
it is possible to eliminate it either by restricting oneself 
to local clusters at $z\lsim0.3$ located in the directions 
of the z=0 CMB quadrupole where the ipCMB vanishes, 
or by using intensity maps of the tSZ effect and a model 
for the temperature distribution of the ICM. We will not tackle this term in detail 
as we consider that  it is well understood and straightforward to simulate.

However, the relative amplitude of the ktSZ and the \ptSZ effects inside and in the vicinity of massive clusters 
calls for a more quantitative assessment as this is expected to depend strongly on the density 
and on the temperature profiles of the cluster. We will analyse the \ptSZ effect at some length, for three reasons. 
Firstly, the degree of CMB polarisation due to this second scattering of the tSZ 
has only been evaluated analytically, using spherically symmetric models for clusters (S99).  
Secondly, as pointed out by \citet[hereafter S80]{Sun80}, it is interesting 
as the second scattering of the tSZ effect  depends only 
on the local free electron density and not on the temperature, 
similarly to the scattering of the intrinsic CMB quadrupole. 
It can therefore be viewed as a means of  detecting large--scale 
structures other than clusters: groups and filaments where 
the inner gas density is a few times the mean cosmic  value. In this regime, 
its key advantage over the polarisation induced by the intrinsic CMB  
anisotropy resides in that it  can be a factor of 10 more efficient, 
depending on observing frequency.  Thirdly, and most importantly,
double scattering can provide information about the inner gas density profiles of clusters, 
cross-checking the constraints derived from direct X-ray observations, which depend 
on a model for the local temperature of the gas when it is not possible to measure it directly.  

Our hydrodynamical simulations show that maps of CMB polarisation can provide 
useful information about the profiles and dynamics of the ICM gas and constrain 
our picture of gas infall, mixture and heating as well as  the dynamics of cold 
subclumps. They indicate a possible way to detect proto-clusters in 
their early stage of collapse. Besides, they show how the cluster  SZ thermal effect can 
polarise the CMB over the filaments of gas between clusters. 
 
The layout of this paper is as follows. In section~\ref{sec:Analytical}, we review the 
analytical formulae we employ to compute the degree and direction of polarisation, 
due to kpSZ effect and to
second scatterings of the tSZ effect, i.e. the \ptSZ effect. We 
present a semi-analytical description of the latter which enables us to obtain a 
substantial gain in computational time. We also summarise our assumptions. 
In section~\ref{sec:Simus}, we describe the setup of the simulations and the 
construction of the polarisation maps.  Section~\ref{sec:Discuss} presents 
the strongest dependences of our results on the cluster gas profile. Section~\ref{sec:Observations}
discusses a general observational methodology to deal with contamination
effects in order to retrieve the original projected density-weighted tangential and radial velocity field.
Conclusions are presented  in section~\ref{sec:CCL}.

\section[]{Analytical description}
\label{sec:Analytical}

In this section, we briefly summarise  the analytical expressions we use for the main terms contributing 
to the polarisation of the CMB induced by the SZ effect discussed by 
S80 (see also S99,  \citealt{Chlu02}). 

We fix an orthonormal frame (\coordpm) at the point $P$ where the
polarisation is induced: it is the position of the scattering electron
in the case of single scattering, or the position of the last
scattering electron in the case of double scattering.  We choose the
\vechmm{Z} direction along the line of sight, from $P$ to the
observer. \vechmm{X} and \vechmm{Y} label the sky coordinates (we
assume an euclidian patch).  We use the Stokes parameters ($I$, $Q$,
$U$, $V$) to describe the amount of polarisation as seen by an
observer looking in a given direction.  $I$ gives the total intensity
of the light, while $Q$ and $U$ measure the degree of linear
polarisation, and $V$ the degree of circular polarisation.  Thomson
scattering of anisotropic but unpolarised incoming radiation cannot
generate circularly polarised radiation so that $V=0$.  The degree of
polarisation is given by:
\begin{equation}
  p = \frac{\sqrt{Q^2+U^2}}{I}, 
\end{equation}
and the direction of linear polarisation $\phi$, measured with respect to the \vechmm{X} axis, is:
\begin{equation}
  \tan (2\phi) = \frac{U}{Q}.
\end{equation}

In the rest of this section we deal with local effects on the line of sight. This approach is valid 
since the Stokes parameters are additive along the line of sight and the resulting $Q$ and $U$ seen by the observer read as follows:
\begin{equation}
  \left\{
    \begin{array}{rcl}
      Q & = & \int~dQ \\
      U & = & \int~dU,
    \end{array}
    \right.
\end{equation}
where $dQ$ and $dU$ are local Stokes parameters at position (\coordpm). As we deal with small optical depths throughout 
the paper we assume that the background intensity $I_{0}$ of the CMB is not affected by SZ scattering:  we set 
$I_{\rmn{tot}} = I \simeq I_0$ to compute the degree of polarisation.

We now present in turn (1) the polarisation induced by the transverse motion of the gas with respect to the  line of sight, 
(2) the effects of double scattering, and (3) the contribution due by scattering the CMB intrinsic quadrupole 
which together generate  most of that  \emph{part} of the CMB polarisation due to the SZ effect. We will mention in the last 
paragraph of section~\ref{sec:Observations}   two other terms which  
 can contribute a large fraction of the \emph{total} polarisation 
of the CMB at a given frequency: the intrinsic polarisation imprinted at last scattering 
and the main types of foregrounds, i.e.  dust grains and point sources emitting synchrotron radiation.

\subsection{Kinetic polarisation}
\label{sec:Analytic:KPol}

The kinetic polarisation kpSZ comes from single scattering of a CMB photon by a free electron moving in the plasma.  
Here, the electron rest frame anisotropy of the radiation necessary for Thomson scattering to
 generate polarisation is entirely due to the electron velocity (S80). We 
refer the reader to S99 and S80 for the detailed derivation 
and we just summarise  their results here.

The $Q$ and $U$ components are given in the rest frame of the observer (all quantities depend on (\coordpm), though not explicitly written):
\begin{equation}
  \begin{array}{rcl}
    dQ & = & - 0.1 I_{0\nu} \sigma_T f(\xnu) n_{e} \beta_t^2 \cos(2\chi)~d\vechm{Z} \\
    dU & = & - 0.1 I_{0\nu} \sigma_T f(\xnu) n_{e} \beta_t^2 \sin(2\chi)~d\vechm{Z} 
  \end{array}
  \label{eq:kinetic}
\end{equation}
Here, $\sigma_T$ is the Thomson cross-section, $n_e$ is the local density of electrons, $\xnu$ is the dimensionless parameter of the Planck law: $\xnu = \frac{h\nu}{kT}$, and:
\begin{equation}
	f(\xnu) = \frac{e^{\xnu}(e^{\xnu}+1)}{2\;(e^{\xnu}-1)^2}{\xnu}^2
  \label{eq:fdef}
\end{equation}
is the spectral shape of the polarisation induced by the CMB in a non-relativistic approximation and
simply is a translation of  the black-body spectrum because of Thomson scattering,  
$\mu= \frac{\vec{v}\cdot\vechm{Z}}{|\vec{v}|}$, where \vecm{v} is the velocity of the electron and
 $\beta_t^2 = \beta^2 (1-\mu^2)$ is its tangential component, and $\chi$ is the angle 
between the tangential velocity and the \vechmm{X}-axis in the tangential plane.
This effect is simple to implement as one simply needs to know the local state of the gas: 
given the value of each field everywhere in the simulation the summation is straightforward.

\subsection{Double scattering due to finite optical depth}
\label{sec:Analytic:Double}

S80 noted that since the optical depth $\tau$ along the line-of-sight through 
massive clusters can reach $\sim0.03$, double scattering cannot be neglected. In this case, 
CMB polarisation is due to second scattering of CMB secondary anisotropies generated by a first scattering 
in a dense region, either in the same structure or in a nearby one. 
The whole process can be alternatively viewed as scattering of the tSZ and kSZ effects. 
However, due to the kSZ effect being typically more than an order of magnitude smaller than the tSZ 
effect for \emph{massive} clusters ( $\gsim 6\times10^{14}$\msunKC and as we will show that scattering of the tSZ (\ptSZ) gives a degree of 
polarisation of the same order as the kpSZ effect of the previous paragraph, it is safe to neglect scattering of the   
kSZ (\pkSZ). S99 give an explicit formal formulation of this second scattering effect:  

\begin{equation}
  \begin{array}{cl}
    dQ = & d\vechm{Z} \frac{3}{16\pi} \sigma_T n_{e} (\coordp) \\
         & \times \int d\Omega~\sin^2(\theta)\cos(2\phi) \Delta I_\nu(\coordp,\theta,\phi) \\
    dU = & d\vechm{Z} \frac{3}{16\pi} \sigma_T n_{e} (\coordp)  \\
         & \times\int d\Omega~\sin^2(\theta)\sin(2\phi) \Delta I_\nu(\coordp,\theta,\phi),
    \end{array} \label{eq:finite_opt_depth}
\end{equation}
where: 

\begin{equation}
 	\Delta I_\nu(\coordp)=y_{\rmn{eff}}(\coordp,\theta,\phi) f_T(\xnu) I_{0\nu},
\end{equation}
\begin{equation}		
	d\Omega = \sin(\theta)\,d\theta\,d\phi,
\end{equation}
$f_T(\xnu)$ is the frequency dependence of the tSZ effect:  
\begin{equation}
	f_T(\xnu) =  \xnu \frac{e^\xnu}{e^\xnu-1} \left(\xnu\frac{e^\xnu+1}{e^\xnu-1}-4\right),	
	\label{eq:fTdef}
\end{equation}
and $I_{0\nu}$ is the intensity of the CMB at 
frequency $\nu$. $\theta$ and $\phi$ give the direction of the  radiation incoming to  the second scattering electron located at ($\coordp$). 
With our axes, $\theta$ is the angle with respect to \vechmm{Z} (colatitude)  while  $\phi$ is the counterclockwise angle to \vechmm{X} in the (\vechmm{X},\vechmm{Y}) plane. 
$y_\rmn{eff}$ is the integral of $\eta\,d\tau$ as seen from ($\coordp$) in the outward direction $(\theta, \phi)$ and is just the 
Compton y parameter in that direction. Interestingly,  S80 pointed out that the degree of polarisation in double scattering at fixed tSZ 
effect ``background'' only depends on the local optical depth, and could be used to probe cooler regions than possible 
with temperature maps of the tSZ effect itself, provided of course that there is already 
sufficient sensitivity to measure the \ptSZ effect in the core regions of galaxy clusters.

In the following, we describe our computation in some length as we show that the \ptSZ effect 
depends sensitively on the detailed gas density profile of clusters, 
that it can in fact dominate in amplitude over the kpSZ effect
and so should not be neglected in single frequency observations, and that it could also 
be of interest as  a means of inducing large-scale polarised CMB radiation.

\subsubsection{A spherical isothermal model for first scatterings}
\label{sec:Analytic:Double:Spherical}

The double integrals in \eqref{eq:finite_opt_depth} are difficult to compute numerically: one needs 
to first evaluate from each point of the simulation volume an integral in all directions, and then 
to perform the summation along the line of sight.  Therefore, we have adopted  
a semi-analytical model: we first choose a spherical, isothermal model 
for the clusters -- here a $\beta$-model -- to be able to compute $y_{eff}(\coordp,\theta,\phi)$ 
for each cluster. We define the $\beta$-model as:
\begin{equation}
  \rho(r) = \rho^{*}\left[{1+\left(\frac{r}{r_c}\right)^{2}}\right]^{-3\beta/2}
\end{equation}
(see, e.g., \citealt{Kom01}) where $r_c$ is the core radius of the cluster and $\rho^{*}$ is a normalisation constant and we take $\beta=2/3$.

The important task is to estimate $\rho^{*}$ and $r_c$ for each cluster of the simulation. The optical depth as seen from 
$(\coordp)$ in the direction $(\theta, \phi)$ can be computed as:
\begin{multline}
  \tau(\coordp, \theta,\phi) = \sigma_T \frac{1}{m_p \mu} \rho^{*} \int_0^{\infty} \frac{1}{1 + \left(\frac{r(l)}{r_c}\right)^2}~dl \\
  = \sigma_T \frac{1}{m_p \mu} \rho^{*} r_c^2 \frac{1}{\sqrt{r_c^2 + r_0^2 - g^2(\coordp, \theta,\phi)}} \left(\frac{\pi}{2} \right. \\
  \left. + \tan^{-1}\left(\frac{g(\coordp, \theta,\phi)}{\sqrt{r_0^2 + r_c^2 - g^2(\coordp, \theta,\phi)}}\right) \right),
  \label{eq:expand_opt_depth}
\end{multline}
where $l$ is the distance from the second scattering electron 
to the point taken along the line of sight  $(\theta, \phi)$, $r_0$ is the distance between the point 
$(\coordp)$ and the cluster located at ($\vechm{X}_c,\vechm{Y}_c,\vechm{Z}_c$), $m_p\, \mu$ 
is the mean particle mass of the ICM, and: 
\begin{multline}
  g(\coordp, \theta, \phi) = (\xp-\vechm{X}_c) \sin(\theta) \cos(\phi) + \\
  (\yp-\vechm{Y}_c) \sin(\theta) \sin(\phi) + (\zp-\vechm{Z}_c) \cos(\theta).
  \label{eq:gdef}
\end{multline}
$y_{\rmn{eff}}(\coordp, \theta,\phi)$ is here the product of $\tau(\coordp, \theta,\phi)$ with the mean relative temperature $\eta=\frac{k_\rmn{B} T_e}{m_e c^2}$, 
where $T_e$ is the mean temperature of the electrons, $m_e$ is the electron mass and $k_\rmn{B}$ is Boltzmann's constant. 
Once we have chosen the set of clusters  which make the most contribution to the tSZ effect over most of the simulation, 
at each point $(\coordp)$ we can add up the local Stokes parameters obtained from~\eqref{eq:finite_opt_depth} and~\eqref{eq:expand_opt_depth} 
for each of these clusters to obtain the local $dQ$, $dU$ due to the \ptSZ effect. 

The underlying assumptions here are first, the validity of the $\beta$-model for the radial profile of the gas, i.e. 
the validity of spherical symmetry and isothermality for the gas given that the X-ray surface 
brightness would be fitted by a $\beta$-profile, second, the validity of isothermality 
in the computation of the tSZ effect itself and third, the fact that lower mass groups 
and filaments will contribute very little to the tSZ effect as 
observed from any point of the simulation.

\subsubsection{Estimating the core radius}
\label{sec:Analytic:Double:Core}

Because we find the relative amplitude of the \ptSZ  to the kpSZ effect to sensitively  
depend on the gas distribution, for massive clusters intersecting along the   line of sight, 
we have computed the normalisation of the $\beta$-model in two ways. This ensures that  
 we are not affected by resolution issues inherent to SPH such as gravitational softening.  
We first estimate the parameters  numerically under 
the assumption that the cluster gas exactly follows a $\beta$-model, and  
we then derive an empirical best-fit to the actual simulated density profile of each cluster. 
This comparison also checks the validity of the $\beta$-profile for the simulated clusters. 

\paragraph*{Numerical approach}

To estimate the free parameters $\rho^{*}$ and $r_c$ numerically, we first identify clusters using a friends-of-friends groupfinder 
\citep{Da85} with linking length 0.2 times the mean interparticle separation. We then compute the virial mass and radius \Mrvir and \Rrvir 
starting from the most bound gas particle of the cluster.  Note that we use only gas particles to estimate \Mrvir and \RrvirDot 
We define the virial radius as the point where the enclosed gas density falls to $\Delta_c\,\rho_c\,f_{\rmn{bar}}$ where $\Delta_c$ 
is the virialization overdensity \citep{Br98}, $\rho_c$ is the critical density of the universe and $f_{\rmn{bar}}=0.13$ 
is the mean cosmological gas fraction. We suppose that the cluster only extends up to the virial radius. 
Starting with an analytical expression for the virial mass \Mrvir as a function of $\rho^{*}$ and $r_c$, 
it is possible to express the mass inside the core radius $m_c(r_c)$ as a function of \Mrvir and $r_c$.
At fixed \MrvirCom we then iterate to find the core radius: it is obtained by choosing a test radius $r_t$, computing the mass $m(<r_t)$ inside $r_t$ and comparing 
it to $m_c(r_t)$. Practically, we start from the most bound particle of the cluster and increase $r_t$, 
up to the desired core radius where $m(r_t)=m_c(r_t)$  where we then set $r_c=r_t$.

\paragraph*{Best-fit approach} 

As a second approach, we directly fit the $\beta$-model to the gas density profile measured in the simulation.  
We employ an adaptive number of points depending on the number of particles in the cluster so that noise 
fluctuations on small haloes can be reduced. We always take 
more than 300 particles per halo to give  a good fit,  with the number of fitting points  less than 100, and 
 we limit ourselves again to the virial radius. We find reasonable agreement between the two approaches, showing that the $\beta$-profile 
and spherical symmetry are fair assumptions. When computing the \ptSZ effect in the simulations,  
we employ the second approach which gives fast results and depends on both the inner and outer parts of the cluster.  
For the massive cluster that we find in our high resolution simulation we compute $\Mrvir=1.5\times 10^{15} \msun$ and the 
 numerical and best-fit approaches give respectively  $r_c=250$~\kpc and $370$~\kpcDot

\subsubsection{Semi-analytical results}
\label{sec:Analytic:Double:SA}

\begin{figure}
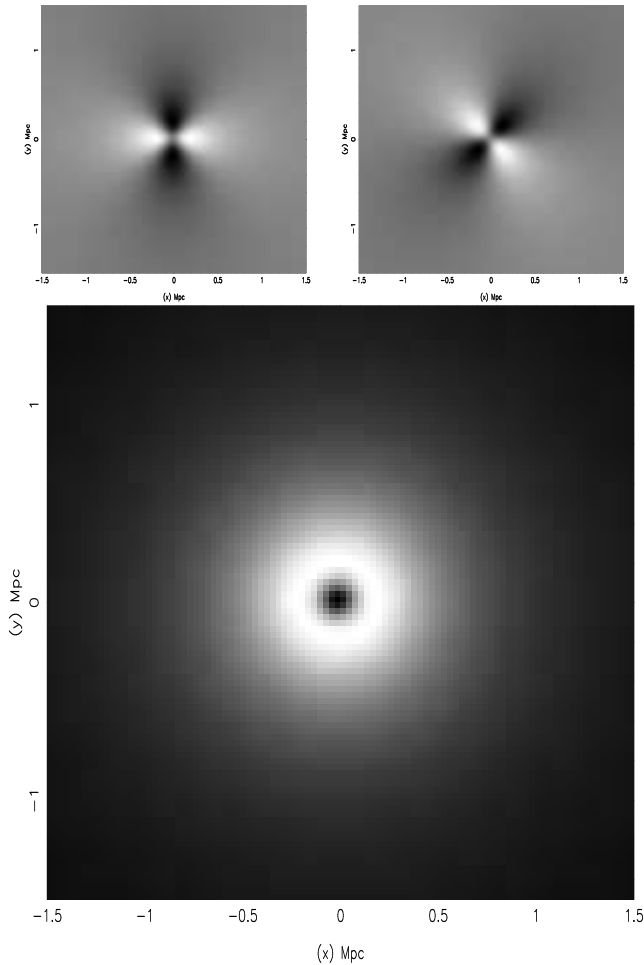

  \begin{center}
    \begin{tabular}{cc}
      \scalebox{1}[1.45]{\includegraphics[angle=270,width=0.45\columnwidth]{figure1a}} &
      \scalebox{1}[1.45]{\includegraphics[angle=270,width=0.45\columnwidth]{figure1b}}
    \end{tabular}
    \scalebox{1}[1.45]{\includegraphics[angle=270,width=\columnwidth]{figure1c}}
  \end{center}        
  \caption{\label{fig:semi_analytic_tau}
    The top images gives the value of $Q$ and $U$ (left and right 
    panels respectively) of the polarisation of the CMB due to second
    scattering of the \emph{in-situ} tSZ effect for a 150 kpc 
core radius cluster put at the centre. The figure illustrates the spatial
variations of $U$ and $Q$ but the real level of polarisation will include
an overall factor dependent on the normalisation of the matter density profile. 
The gray scale is linear in $Q$, $U$ and the polarisation. Black is for the minimum relative value and white for the maximum relative value. The change in the intensity
    of the polarisation is clear at the centre. As expected the axis of $Q$ and $U$
    are rotated by $\pi/4$ and the axis of $Q$ is orthogonal to the $x$-axis.
    The global degree of the polarisation, as given by $\sqrt{Q^2+U^2}$ for $I=1$,
    is shown in the lower image. Note the annulus of high intensity 
    around the core and the fall-off at the centre as spherical symmetry
    washes out the signal.}  
\end{figure}

Having set the gas density profile for the integration of~\eqref{eq:expand_opt_depth} over the path of the incident radiation 
on the second scattering photon, we proceed to integrate ~\eqref{eq:finite_opt_depth} over $(\theta,\phi)$ numerically. 
Figure~\ref{fig:semi_analytic_tau} shows the results of the computation of $Q$ and $U$, 
for a spherically symmetric gas distribution  following a $\beta$-profile with core radius $150$~\kpcDot
Here and in all subsequent figures we take $x_{\nu}=3$: we discuss the frequency dependence of the 
effects in section~\ref{sec:Discuss}.

As expected, the polarisation has circular symmetry. The signal vanishes at the centre of the cluster because of 
the particular symmetry involved at that point: the centre of the cluster does not 
see any changes due to  any rotations of the space around it, so that there can be no polarisation induced at this point. 
As a consequence, the intensity pattern shows an annulus: we checked that the radial 
distance $r_{\tx{max}}$ to the circle on the annulus where 
the degree of \ptSZ is maximum scales as $r_{\tx{max}}=1.3\, r_c$.  Note that this 
feature has already been found by S99 (see their figure 3) who also integrate their equations numerically:  
their King profile with n=1 is similar to the $\beta$-profile assumed here.

\subsection{Polarisation due to the CMB quadrupole}
\label{sec:Analytic:Quad}

Thomson scattering of the intrinsic quadrupole of the CMB by the free electrons of the ICM will 
contribute to the degree of polarisation observed toward massive clusters. S99 show 
that this term can reach $2\times 10^{-6} \; \tau \; \sim \;4\times 10^{-8} $ for $\tau=1/50$, 
comparable and even higher than the polarisation due to the kpSZ or \ptSZ effects 
presented above. However, at low redshift, there are  four orthogonal directions in the sky 
where one expects the degree of CMB polarisation 
due to the scattering of the quadrupole component to vanish (see Fig.1 of S99).
In practice, the directions towards which the CMB quadrupole vanishes have been determined 
with some accuracy by \emph{WMAP}: one is near $(l,b)\simeq (-80\degr, 60\degr)$, a direction close 
to the Virgo cluster \citep{Tegmark2003}. Note that this only holds for local clusters ($z\lsim0.3$), 
as objects at higher redshift will observe a different realisation of the 
primordial quadrupole \citep{Kam97}.

\section{Rendering a galaxy cluster}
\label{sec:Simus}

\subsection{Parameters and setup}
\label{sec:Simus:Pars}

We use the public version of  {\sc gadget}~\citep{SprGadget2001} to run the simulations after  
modifications to include  the ``standard entropy'' formulation \citep{SprEntropy02}, cooling of the gas and photoionisation by a 
uniform UV background.  The simulations  follow a cube of $70$~\Mpc comoving side and assume  
a concordance $\Lambda$CDM cosmology with $\Omega_0 = 0.3$, $\Omega_b\:h^{2} = 0.019$, $\Lambda = 0.7$, $h = 0.7$. 
We have not switched on cooling and photoionisation  in this study but we will analyse their impact in a forthcoming paper ; 
here the evolution is fully adiabatic which suffices  for our purpose. 

We constrain the density field using the implementation of the 
Hoffman-Ribak algorithm proposed by \citet{Wey96} to obtain 
a massive cluster at the end the simulation. 
To obtain a cluster of approximately $1.5\times10^{15}$\msun we constrain  the initial conditions to have a gaussian  peak  with $\sigma=13.5$~\Mpc 
in the centre of the box. 
The height of the peak is three times the rms value obtained by smoothing an unconstrained 
density field with the same gaussian kernel.  At z=0, the optical depth of the central region of the cluster 
is  $\sim3 \; 10^{-2}$ and the mean electron temperature in the virialised gas is $\sim15$~keV: the cluster is more massive than Coma, and
probably closer in mass to Perseus.

To speed up the simulation we have used mesh refinement when constructing the initial conditions: 
after  running a low resolution simulation the particles of the cluster 
have been traced back to the initial redshift  to find the  Lagrangian region containing all the mass virialised by z=0. 
We have used two simulations: in the first the initial density field is realised on a $64^3$ mesh embedded in a $32^3$ mesh covering the whole volume, 
while the second uses a $128^3$ mesh embedded in a $64^3$ mesh. In the two simulations, we have used the same phases 
for the common density perturbation modes so as to verify the stability of the method with respect to mesh resolution. We put high mass 
DM and gas particles on the nodes of the low-resolution grid and low mass DM and gas particles on the nodes of the high resolution grid.  
Following common practice, we set the softening length of both low and high resolution particles 
(we use the subscripts ``lr'' and ``hr'' in the following) to one tenth of the their mean interparticle separation. 
The softening length $l_{\rmn{soft}}$ is kept fixed in comoving coordinates:  $l_{\rmn{soft,lr}}=400\kpc$ and $l_{\rmn{soft,hr}}=30\kpcDot$ 
We have checked that the overall intensity of the effects and the gas distribution is the same within a factor of two between the simulations, 
and in the rest of the paper we will only discuss the simulation with high mesh resolution. 

\begin{figure}
  \scalebox{1}[1.45]{\includegraphics[angle=270,width=\columnwidth]{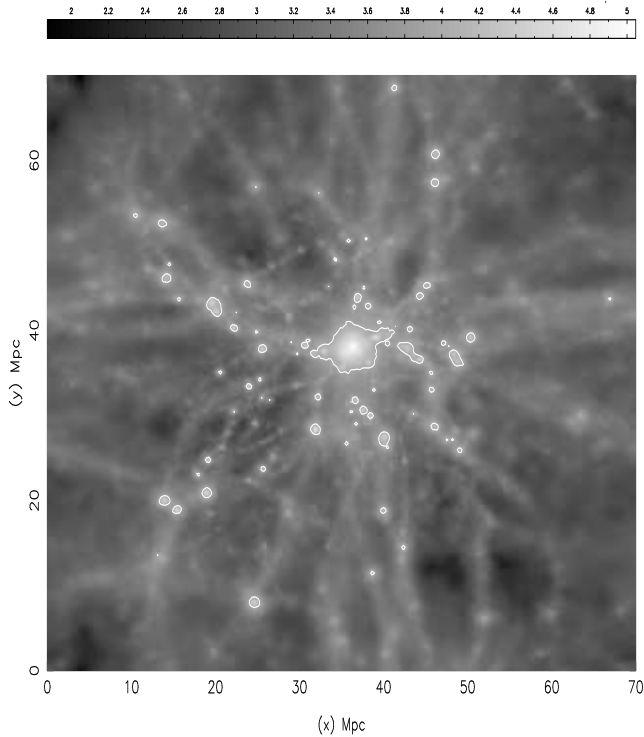}}
  \caption{\label{fig:density_map}Map of the gas density of the whole high resolution simulation projected along  the line of sight. 
    The side is $70$ Mpc and the cluster is located close to the middle of the box (as it is along the line of sight). 
    There are four filaments reaching the cluster and extending at close to right angles. 
    The colour scale is in decimal logarithm of arbitrary units in projected density.} 
\end{figure}

The z=0 map of the density projected along the \vechmm{Z} direction of the whole box  is shown  in Figure~\ref{fig:density_map}. The box is 70 \Mpc wide. 
The friends-of-friends groupfinder finds a little more than $3.5 \times 10^5$ DM particles in the central cluster, while there are 
$N_{\rmn{lr}}=171019$ and $N_{\rmn{hr}}=729000$ DM particles in total, with masses $M_{\rmn{lr,DM}}=2.61\times 10^{11}$\msun and $M_{\rmn{hr,DM}}=3.26\times 10^{10}$\msun 
and the same number of gas particles with masses $M_{\rmn{lr,gas}}=3.93\times 10^{10}$\msun and $M_{\rmn{hr,gas}}=4.92\times 10^{9}$\msunDot

\subsection{Mapping the SZ effect}
\label{sec:Simus:Pics}

\begin{figure}
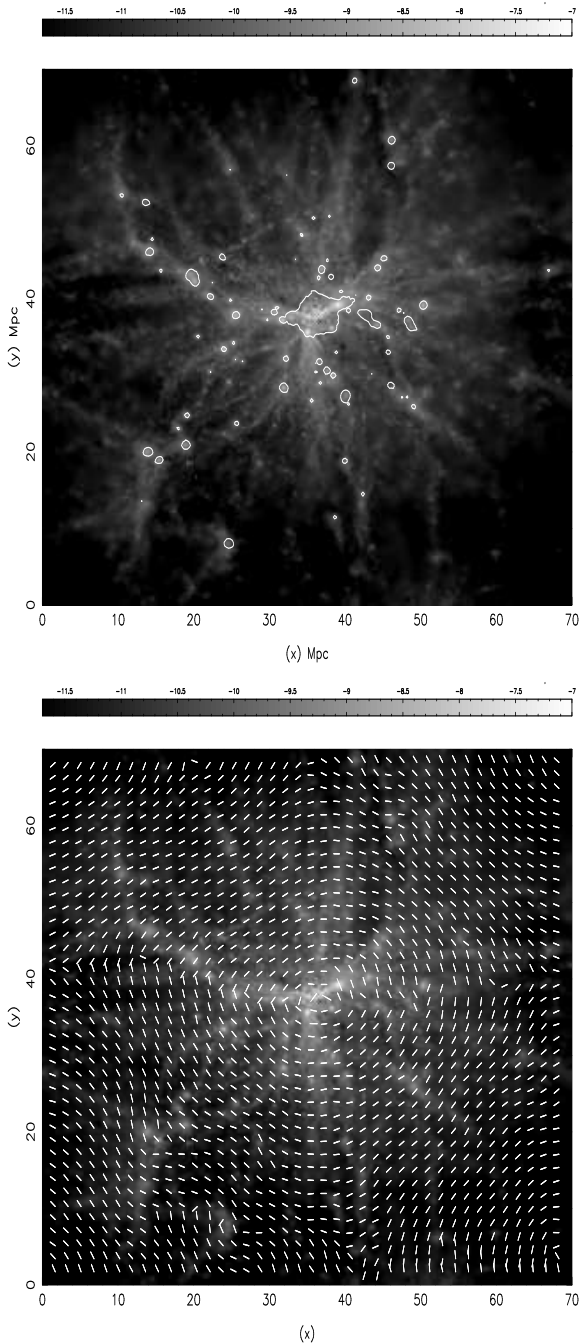

  \begin{tabular}{c}
    \scalebox{1}[1.45]{\includegraphics[angle=270,width=0.9\columnwidth]{figure3a}} \\
    \scalebox{1}[1.45]{\includegraphics[angle=270,width=0.9\columnwidth]{figure3b}}
  \end{tabular}
  \caption{\label{fig:thermal}Map of the \ptSZ effect for the main cluster of the highest resolution simulation. 
The figure shows a box  $70$~\Mpc wide and the whole box has been projected. 
The decimal logarithm of the degree of polarisation is colour-coded with the scale provided at the top. 
This scale has been frozen to facilitate  the comparison with the other maps. Radial lines coming from the 
central cluster are an artifact of the limited number of steps for the integration on $\phi$ in equation \eqref{eq:finite_opt_depth}; they 
could be avoided but at the expense of unreasonable computing time, the CPU scaling as the number of iterations. 
Note the fall-off of the \ptSZ effect as one approaches the centre of the cluster: 
increasing symmetry tends to wash out the signal. 
Note also the residual polarisation on the filament leaving to the east: 
this shows the possibility of the \ptSZ effect  tracing  large-scale structure.}
\end{figure}

\begin{figure}
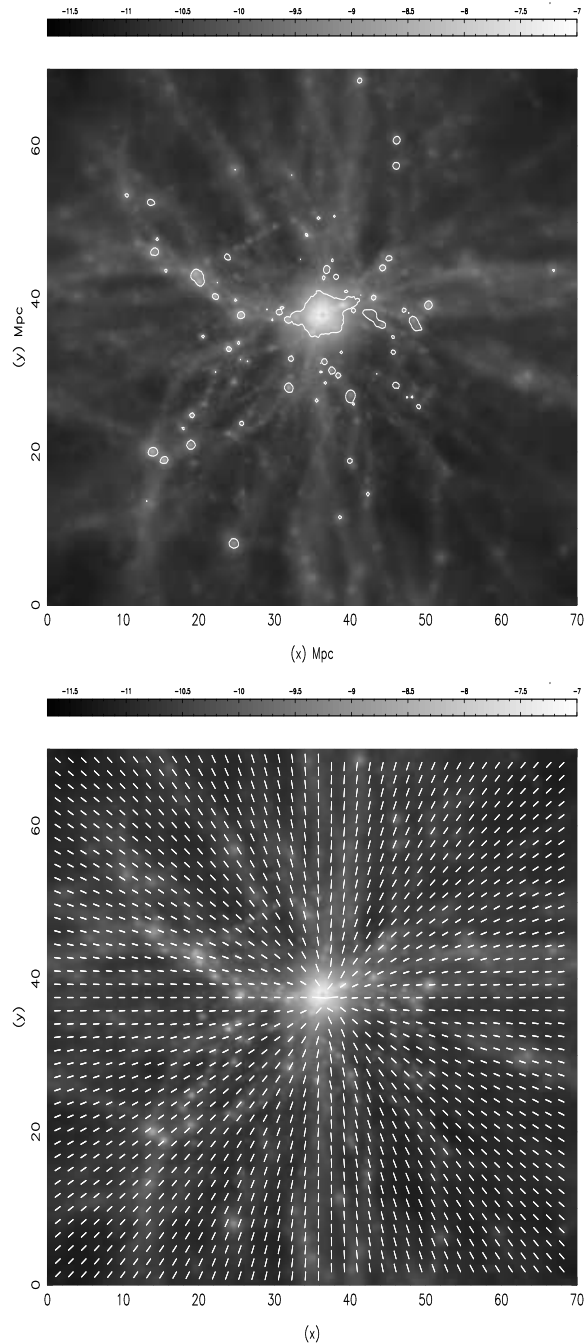

  \begin{tabular}{c}
    \scalebox{1}[1.45]{\includegraphics[angle=270,width=0.9\columnwidth]{figure4a}} \\
    \scalebox{1}[1.45]{\includegraphics[angle=270,width=0.9\columnwidth]{figure4b}}
  \end{tabular}
  
  \caption{\label{fig:kinetic}Maps of the kpSZ effect in the highest resolution simulation. The figure shows a box  $70$~\Mpc wide and the 
whole box has been projected.  The decimal logarithm of the degree of polarisation is colour-coded with 
the scale provided at the top (it is the same scale as in Fig.~\ref{fig:thermal}). It has been computed using equation \eqref{eq:kinetic}. 
Contours of the projected gas density have been added to the top image 
 to make a precise comparison with previous maps. 
The directions of polarisation due to the kpSZ effect are shown on the bottom image.} 
\end{figure}

\begin{figure}
  \scalebox{1}[1.45]{\includegraphics[angle=270,width=\columnwidth]{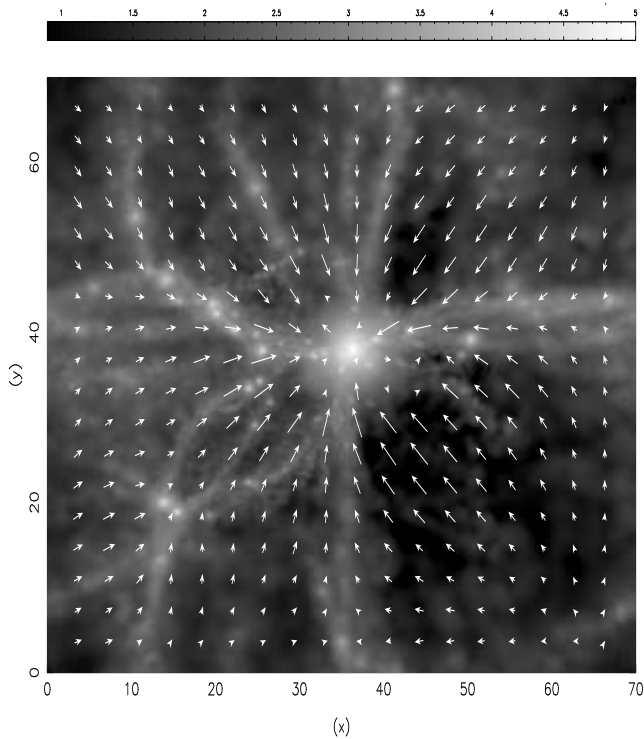}}
  \caption{\label{fig:velocities}Maps of the projected density and of the projected gas velocity field in 
the highest density slice of the simulation. The width is 70 \Mpc and the thickness 3 \MpcDot
The length of the arrows is proportional to the velocity with  a maximum of approximately $2000$~\kms.} 
\end{figure}

\begin{figure}
  \begin{center}
    \scalebox{1}[1.45]{\includegraphics[angle=270,width=0.9\columnwidth]{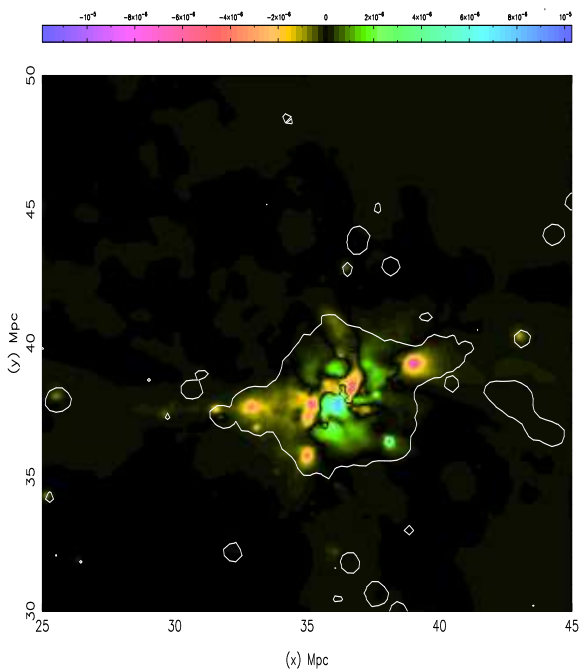}}
  \end{center}
  \caption{\label{fig:velocity_rad}Map of the kSZ effect in terms of $\Delta T/T_{CMB}$. The change in temperature is shown 
with a colour \emph{linear} between $\simeq -10^{-5}$ and $\simeq 10^{-5}$. Density contours 
have also been overplotted on the map but please note that it is now only $20\;\Mpc$ wide. Small scale details have been smoothed out as we 
have the convolved the image with a Gaussian beam of FWHM 1'}
\end{figure}

To render the SZ effect we have built a parallel imager capable of integrating any physical quantity along the line of sight in the simulation. 
The imager includes the \sph kernel of {\sc gadget}~and tree neighbour searching so that physical quantities are well 
computed at each 3D pixel of the simulation volume. The tree is filled with the \sph particles and  we use this to compute the local physical effects 
with the same algorithm  that is used to  estimate hydrodynamical forces on a particle. We then integrate up the results along the line of sight (here the \vechmm{Z} direction).
Figures~\ref{fig:thermal} and~\ref{fig:kinetic} show the \ptSZ effect and the kpSZ effect respectively. The figures show a box  70 \Mpc wide 
and we have projected the full thickness of the simulation.  Furthermore, to compare 
with the distribution of the projected gas velocities we plot these in Figure~\ref{fig:velocities}, where 
we have projected only the gas velocities and  gas densities from a 3 \Mpc thick slice  
containing the highest density region of the simulation i.e. the centre of the cluster (the width is again 70 Mpc).  We
select isodensity levels in Figure~\ref{fig:density_map} and show them on the other figures to make a comparison 
between the polarisation signal and the possible presence of clumps of dense gas. The colour scales for the polarisation 
plots have been kept the same to facilitate the comparison.  Following \citet{Nag03b} we have tested the effect of convolving the 
resulting maps with a gaussian kernel corresponding to  $1$ arcminute FWHM. Putting the cluster at z=0.1 this would  
 represent a physical scale of 77.5 \hkpcDot In practice, we find that convolution leaves the large scale features of the maps unchanged but 
removes some details in the zoomed map of the kSZ effect shown in Fig.~\ref{fig:velocity_rad} which has a 20 \Mpc width. As a result, 
we present all maps but that of Fig.~\ref{fig:velocity_rad} unconvolved for simplicity. 

\section{Dependence on physical parameters}
\label{sec:Discuss}

In this section, we first discuss features of the \ptSZ effect 
associated with resolution issues in the core of the cluster. 
Then we compare the \ptSZ effect to the kpSZ effect. The 
location and intensity of the bright annulus of the \ptSZ effect 
can provide useful cross-checking information on X-ray observations and in particular 
on the validity of the $\beta$-profile as an accurate description of the 
projected inner density of the gas, while the kpSZ effect can constrain the projected density-weighted tangential velocity of the gas.

\subsection{ Dependence of the \ptSZ effect on cluster internal gas profile}
\label{sec:Discuss:internal_structure}

\subsubsection{Effect of the internal structure on the intensity of the \ptSZ effect}

As we have pointed out above, the computation of the polarisation due to 
double scattering involves two integrals. The first one integrates the effect along 
the line of sight while the other calculates the anisotropy of the incoming radiation  
 at each point of the line of sight.  In regions of high electron density, the two 
consecutive scatterings will happen at almost the same location and we should therefore expect 
the polarisation term due to double scattering to contain a strong dependence on the 
 spatial distribution of electrons in the cluster. In fact, if we calculate 
the Stokes parameters for a toy model (here a $\beta$-model for the gas density profile), 
we find that the overall signal will scale as $n_{0}^2 r_{\rmn{c}}^4$.  From this example 
we infer that the double scattering term will be very sensitive to the internal  
gas structure of the cluster like the inner radial density slope 
or the possible presence of a core which will affect both $r_{\rmn{c}}$ and $n_{\rmn{0}}$.  If the 
angular resolution is sufficient, the \ptSZ effect could detect clumps of 
dense cold gas orbiting in the ICM such as remnants  of minor mergers or even 
cold fronts provided their density contrast with respect to the surrounding hot ICM is high enough.

We have already noted in paragraph~\ref{sec:Analytic:Double:SA} 
that the two scattering terms will show a very particular polarisation pattern with an annulus around the centre of 
the cluster, and that the radius $r_\tx{max}$ of the maximum of intensity of this annulus  is simply related to the 
core radius if one assumes a $\beta$-profile.

\subsubsection{Effect of resolution on the estimation of the core radius}
\label{sec:Discuss:Core}

As pointed out above, the maximum degree of polarisation  $p_\rmn{max}$ of
 the radiation after the second scattering of the thermal effect depends
 significantly on the core radius of the $\beta$-profile of the gas. Consequently, 
$p_{\rmn{max}}$ may easily vary by a factor 2 for reasonable
 values of $r_{\rmn{c}}/\Rrvir$. This can be advantageous in the case of multi-frequency
 observations, where one can separate the kpSZ from the \ptSZ effect and use
 only the latter to constrain the gas density in the central region. On the 
other hand, it shows that it will be difficult to disentangle both effects
 when only a single frequency is available.  In any case, it is essential to
 have a robust  estimate of the order of magnitude of the  intensity of the polarisation signal
which is not plagued by resolution issues. In all the above,  we have ignored the central softened part of the 
cluster as we computed the core radius of the gas according to the $\beta$-model
because it introduces artificial smoothness in the derived cluster gas profile. 
We have noticed that when we skip the inner region as one would do to suppress 
the dependence on the resolution of the simulations,  the order of magnitude 
of the \ptSZ effect can significantly increase  if  one uses 
the ``numerical approach'' of~\ref{sec:Analytic:Double:Core} to estimate it.  
However, the cluster gas profile parameters $r_{\rmn{c}}$ and $\rho_{*}$ obtained 
in this case with the numerical approach are reasonably close to those resulting 
from best fitting a $\beta$-model, showing that the amplitude we 
derive for the \ptSZ effect inside massive clusters is robust in this case.  

To summarise, even if a very precise computation of the intensity of the \ptSZ
inside clusters will require accurate knowledge of the gas profile, our simulations conversely prove 
that, once it is singled out in observations, it can prove a powerful tool for  constraining the gas profile. 

\subsection{Comparison of the two intensities}
\label{sec:Discuss:PolIntensity}

We have confirmed  numerically the analytical predictions of S80 
and S99 that finite optical depth effects are not negligible and 
that second scattering of the thermal SZ effect may be 
the dominant source of SZ polarisation other than the intrinsic CMB quadrupole 
in the case where the optical depth is high enough. 
In this paragraph, we compare the intensities of the \ptSZ and the kpSZ effects first with respect 
to their spatial distribution in the cluster, then with respect to the observing frequency. 
We also discuss the directions of polarisation of the \ptSZ and the kpSZ effects.

\subsubsection{Spatial variation}
\label{sec:Discuss:Comparison:Spatial}

The peak of the kpSZ effect is located near the centres of gas clumps or close to the centres of clusters but 
seldom directly on them.  As the kinetic polarisation reflects mostly the part of the tangential velocity of the gas 
which is coherent over some scale this is not unexpected because the gas inside discrete clumps 
or in the core region of the cluster may have thermal velocities  largely in  excess of the bulk motions 
while gas accretion on these structures can reach bulk velocities of order of $1000\;\kms$ which does more than compensate for  the difference in projected density. 

When the accreting gas reaches the local density maxima its velocity distribution becomes isotropic, hence reducing the signal.  
This is the case, for instance,  for the infall one can observe just under the central cluster (in green), at $x\sim35$~\Mpc and $y\sim35$~\Mpc
in the bottom map of Fig.~\ref{fig:kinetic}: it is associated neither with a local peak of density nor with the virial radius of the cluster. It 
 is plain gas accretion, in other words, a coherent inflow. On the opposite side of the cluster, 
a clump above the central regions (clearly seen in Fig.~\ref{fig:density_map} 
at $x\sim45$ and $y\sim55$ Mpc) is almost absent from the kSZ map. 
If the gas density profile in the very central region of a halo is sufficiently symmetric, any similar steep variation 
in the tangential velocity of the gas (accretion or outflow because of AGN or SN feedback) 
would lead to possible detection of the inner ring in the kpSZ maps. Farther out, dense cold fronts 
resulting from recent mergers could also make a strong signature in kpSZ maps, 
as they are thought to be contact interfaces separating the hot gas of the ICM 
and the cold gas of the subclumps as the two phases move with transsonic 
velocity with respect to each other.    

It may be noticed that the intensity map for the kpSZ effect is sharper than for the \ptSZ effect. 
This is mainly because the polarisation follows the locations where the gas 
accretes on the high density regions or reaches high velocities, and because changes in the projected velocity field 
of the ICM can be very abrupt and combine with variations of the projected gas density. 
On the other hand, the \ptSZ effect mostly follows the projected 
density and temperature with the latter being much smoother than the variation of velocities.

\subsubsection{Frequency variation}
\label{sec:Discuss:Comparison:Frequency}

The different frequency-dependence of the \ptSZ and kpSZ effects 
will introduce changes in the degree and direction of polarisation 
of the resulting combined signal as the observing frequency changes. 

As an example, we briefly discuss the effects expected 
as the observation frequency changes from 
$\xnu<3.83$ to $\xnu>3.83$, looking at infalling gas 
on the outskirts ($r \lsim r_{\rmn{vir}}$) of a massive cluster. 

For  $\xnu<3.83$ the direction of polarisation of the \ptSZ effect is 
approximately orthogonal to the direction of the kpSZ 
effect. As the cluster is the geometrical centre for both the accretion and the tSZ 
effect, it can be viewed as a point source far from the core. Therefore both effects 
share the same symmetry but one is rotated $90\degr$ with respect to the other. 
Because they have similar intensities they may counterbalance one another 
and  the resulting degree of polarisation can vanish. 
Conversely, if $\xnu>3.83$ the two effects are coherent 
and oriented in the same way so they will amplify: the signal can be twice as large. 
(Recall that in all our figures we have assumed an $x_{\nu}=3$.)

\subsubsection{On the direction of the resulting polarisation}
\label{sec:Discuss:Comparison:PolDirection}

The direction of polarisation of the kpSZ effect is clearly orthogonal to the direction of the tangential velocity of the gas. 
In our simulations the cluster is massive enough to drag the gas from even the distant edges of the box: 
on the velocity fields, arrows in the outskirts of the box are pointing towards the central cluster. It is interesting to note that along major 
structures like filaments this favoured orientation is also significantly enhanced: coherent motion of the gas is amplified over such regions. 
Conversely, at  this level of resolution,  the orientation of polarisation may point toward a completely arbitrary direction over the close outskirts 
of the cluster: the signal generated by gas accretion can interfere with that due to the presence of slowly moving, high density clumps around the cluster.

\section{Observational prospects}
\label{sec:Observations}

In this section, we go beyond using SZ polarisation to constrain only the dynamics and density profile of the ICM. 
We first discuss how the detection of the kpSZ polarisation could be used as a tracer of collapsing objects. 
Then, we present the \ptSZ effect as a means of  probing external structures. Finally, we discuss the 
contaminating polarisation induced by other dominant SZ effects, by the CMB polarisation, 
by dust, and by the cluster magnetic field.

\subsection{The kinetic polarisation term as a probe of 
collapsing structures}
\label{sec:Observations:Collapse}

The key feature of the polarisation due to the 
kinetic term is its dependence on the tangential component of 
the velocity. In the situation where a structure is collapsing 
under the effect of its own gravity, the velocity vectors point  
toward the centre of the structure. 
In this situation it is possible to calculate the integrated effect 
along the line of sight at the angular direction $\theta$ from the 
centre. The degree of polarisation scales as: 
\begin{equation}
\rmn{p} \propto \int n_e (v \cos(\alpha))^2 dl
\end{equation}
where $v \cos(\alpha)$ gives the tangential component of the 
velocity vector. We can substitute $\cos^2(\alpha) = d^2/(d^2+l^2)$ 
where $d = \theta D_{\tx{ang}}$ with $D_{\tx{ang}}$ 
the angular diameter distance to the cluster.
This integral over the line of sight of the tangential component 
times electron density will reach a maximum at some point between the 
centre of the structure and its radius but this maximum will never 
happen at the centre of the structure since in this case the 
tangential component of the velocity is very small 
($\cos(\alpha) \rightarrow 0$). 
 
\begin{figure}
  \begin{center}
    \includegraphics[width=0.9\columnwidth]{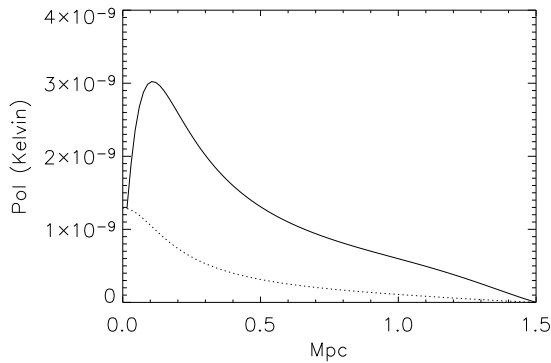}
  \end{center}
  \caption{\label{fig:collapse}
           The solid line shows the polarisation due to the collapse 
           of a cluster ($r_c = 150$ kpc and $n_e=10^{-3} \mbox{gr/cm$^{3}$}$)
           with a homogeneous collapsing velocity of 800 km/s. 
           The dotted line shows the effect due to the bulk motion of 
           the cluster with respect to the Hubble flow for a tangential 
           velocity of 300 km/s. The real polarisation amplitude 
           is the product of the plot times the corresponding 
           frequency dependence.}
\end{figure}

Although the above discussion was focused on a cluster, the same discussion 
will be valid for a structure which is still collapsing, such as 
proto-clusters. This opens the exciting possibility of 
detecting the clusters even before their collapse time.
 This is because the collapse polarisation term is proportional to the 
collapse velocity squared times the optical depth which is the same for a 
cluster already formed and for its progenitor in the collapsing process, 
and also because temperature does not play any role. 
In fact, we only require that the gas cloud is ionised and collapsing. 
It could be possible to see this gas in SZ polarisation forming a ring around the 
centre of the cloud before it starts to emit X-rays 
or before its temperature is high enough to create a substantial thermal 
SZ effect. At distances $r > r_c$, the collapsing term will dominate 
the effect due to the bulk motion of the cluster. This comes from the 
higher collapsing velocities expected for a typical cluster when compared with 
its expected bulk motion. An example is shown in Fig.~\ref{fig:collapse}, where the shape of the kpSZ 
effect expected for a spherical proto-cluster collapsing with a 
homogeneous collapsing velocity of 800 km/s (solid line) is compared to the effect 
of bulk motion of the cluster assuming a tangential peculiar velocity of 300 km/s (dotted line).


\subsection{The finite optical depth effect as a polarisation generator}
\label{sec:Observations:PolSource}

S80 have noted that CMB polarisation due to double 
scattering effects can bring out gas structures much less overdense than the 
virialised regions of massive clusters: smaller groups and filaments 
with sufficient optical depth may become sources of polarised sub-mm radiation, 
although to a very small degree. This is because the second scattering of the primary CMB modified 
by either the tSZ or the kSZ effect does not depend on the local temperature or on the local velocity, 
but only on the local optical depth. As a result, if it is possible to measure the few nK of the \ptSZ 
effect up to the outskirts (e.g., $\Rrvir$) of massive clusters,  
it will be feasible to detect filaments as clusters would provide the tSZ input to generate 
the additional CMB anisotropy. An example of this effect is the left filament 
leaving  the cluster in the NW direction in Fig.~\ref{fig:thermal} (see at $x\sim20$ Mpc and $y\sim40$ Mpc), 
which is less apparent  in the kpSZ map (Fig.~\ref{fig:kinetic}).  This is in contrast to 
the mapping potential  of the tSZ effect in temperature 
which is essentially restricted to lines of sight where 
the integrated Compton parameter $y$ reaches substantial values because at each point, 
the high local temperature must contribute to significant  pressure. 

To summarize, it is possible to detect filaments, first, by mapping an area of the 
sky and second, by modelling the distribution of nearby clusters: it may be possible to predict 
the direction of polarisation expected on a given region of the filament. Here, simulations including 
photoionisation from the UV background that strongly affects the IGM 
are necessary to make a quantitative assessment.

\subsection{Contamination effects}
\label{sec:Observations:Contaminations}

The kpSZ and \ptSZ effects are probes that follow respectively the dynamics and the inner density profile  
of the ICM but it is not clear whether they may be detectable in the near future.  
Intrinsic CMB polarisation, SZ effects (scattering of the intrinsic quadrupole, and possibly  even octupole: \citealt{Challinor2000}), 
the tangential kinetic effect, and second scattering of the tSZ and kSZ effects), foregrounds 
such as Galactic dust and Galactic synchrotron emission or radio point sources with synchrotron emission 
all contribute to the polarisation of the incoming sub-mm signal. These effects 
have a different frequency dependence except for scattering 
of the CMB intrinsic quadrupole which depends linearly on $\xnu$ in the 
same way as does second scattering of the kSZ effect. The 
obvious way to separate components will be with multi-frequency observations. 

However, the situation could be tractable even at a single frequency, provided one obtains 
an image with sufficiently high resolution to avoid point sources. Galactic 
dust and Galactic synchrotron emission induce a high degree of polarisation, but dust 
maps will enable one to correct for some of the contamination or at least to 
set a lower galactic latitude for the observations. 
Magnetic fields in the Galactic ISM will Faraday rotate the plane of polarisation, however,  
Faraday rotation depends strongly on the frequency. 
Because our present knowledge of the Galactic magnetic field is still crude this constraint 
may also translate into a minimum Galactic latitude for observations or to a 
lower bound for the operating frequency.

Removal of the intrinsic CMB polarisation is feasible, as their fluctuations 
are typically on scales larger than the ones considered here 
(i.e. greater than a few arcminutes). An appropriate filtering of the maps can safely 
remove most of this signal while keeping the small scale polarisation in clusters. 
At low redshift when the quadrupole remains close to the one observed at z=0, 
the polarisation component in galaxy clusters due to the intrinsic CMB quadrupole will 
vanish in four directions in the sky which have been measured with some accuracy by \emph{WMAP}. 
 Furthermore, a measurement of the CMB 
quadrupole would in principle enable a modelling of its induced polarisation  
signal in galaxy clusters which could be removed from the data. 

Note that although complex large scale or 
intracluster magnetic fields may again Faraday rotate the plane of polarisation, the rotation angle at 
frequencies of $\nu \sim 300$~GHz is negligible: according to \citet{Oh03}, 
the angular rotation of the CMB quadrupole is typically of order 
$\Delta_{RM} \simeq 1-10\degr(10 GHz/\nu)^2$ on the line of sight through 
a massive cluster. 

Another possible source for generating the CMB anisotropy 
necessary to obtain polarisation inside clusters is through gravitational lensing effects \citep{Gib97}.  
For the fairly massive Abell cluster A576 ($M_{\tx{vir}}\sim 1.1 10^{15}$ \hmsun according to \citealt{Gir98a}),
 \citep{Gib97} finds that the expected contribution from gravitational lensing to the CMB 
polarization is more than an order of magnitude less than the 
expected contribution from the \ptSZ effect. Note that \citep{Gib97} also considered two other gravitational   
effects due to the gravitational bound or to the possible gravitational contraction  
or expansion of the cluster, and found that for A576 both effects can 
be two orders of magnitude greater than the \ptSZ effect. However, 
these estimations have been obtained in the frame of a ``two-steps'' vacuole model \citep{Not84} 
which is not supported by simulations of large-scale structure. This goes  
beyond the scope of the present paper, but it seems necessary to 
quantify the impact of the last two gravitational effects more precisely 
in future work using simulations. 

With multi-frequency observations and extensive contamination removal, we conclude that the kpSZ 
effect can be used to measure the projected, density-weighted tangential velocity of the ICM  in massive clusters, 
to the level shown in Fig.~\ref{fig:kinetic}. On the other hand, the kSZ effect can be used to measure 
the line of sight component of the projected, density-weighted radial velocity of the gas. An example 
of the kSZ effect is shown in Fig.~\ref{fig:velocity_rad}, 
which is a $20\;\Mpc$ wide, 10 kpc thick slice of the central cluster.  
It is obviously possible to compute the velocity of major clumps on the map.
The issue is whether  polarimeters can reach a sensitivity of order of a few nK.

\section{Conclusions}
\label{sec:CCL}

We have carried out high resolution hydrodynamical  simulations of the formation of a massive, Perseus-size cluster to check whether the  polarisation 
of the CMB induced by the SZ effect at $z\sim0$ can provide information about the dynamics and inner density profile of the ICM. 
The contribution to the observed polarisation due to the SZ effect can be split into 
4 major contributions: scattering of the intrinsic quadrupole, 
scattering of the velocity-induced quadrupole and second scattering of respectively, the thermal and kinetic SZ effect. 
The different frequency dependences will enable them to be  separated using multi-frequency observations.  
We have not discussed here the polarisation due to the CMB quadrupole  as it is straightforward to compute, 
 only depending on the projected electron density, and  can also be avoided in four directions of the sky, 
although we note that it can provide  cross-checking information on the gas distribution inside the cluster.  
We have focused on (1) the polarisation due to the velocity-induced quadrupole and on (2)  the polarisation due to 
second scattering of the tSZ effect because they are  possible strong constraints 
on respectively the projected density-weighted tangential velocity of the gas and on 
the projected density of the gas within the cluster core radius, and we have  
shown that  our numerical simulations can make accurate estimates of both amplitudes.

When the kpSZ effect is combined with the kSZ effect, which probes the projected, density-weighted radial velocity of the gas, it is possible 
to deduce information about the 3D dynamics of the ICM, at least in regions where the projection along the line of sight washes 
out only a limited fraction of the signal. This is the case, for instance, at the projected location of dense knots of cold gas 
orbiting the ICM and which reside inside the virial radius as a result of past mergers. 
We plan to tackle how well a typical 3D velocity field can be recovered from both the projected kSZ and kpSZ 
information in a subsequent paper. 

Interestingly, the kpSZ effect  can also prove very useful for 
mapping collapsing structures, where the 
tangential component of the collapse velocity can reach values higher 
than the bulk velocity induced by large-scale structures. This is further facilitated 
as the kpSZ term does not depend on the temperature of the gas and the  
collapsing gas is typically not hot enough to be studied with X-ray observatories until structures are well advanced 
in their formation process. Note also that the optical depth of a collapsing structure is 
redshift independent. As a result, the kpSZ effect allows 
to map these proto-clusters higher in redshift than do other, 
temperature dependent effects.

Although the computation of the polarisation (2) due to second scattering of the tSZ effect is intricate, we have confirmed 
with more detailed modelling, and for the first time in a realistic cosmological setting, 
that this effect can reach the level expected from the scattering of the quadrupole 
induced by the tangential motion of the gas. This \ptSZ effect scatters the first anisotropies 
created by the thermal SZ effect occurring inside massive clusters, and because the tSZ 
anisotropy can be seen much further out without diminution, the resulting polarisation signal can trace relatively 
well structures like groups, filaments and the cosmic web which are much cooler and also less overdense.
 
The challenge is now for CMB polarisation-sensitive instruments to reach the 10 nK 
level typically predicted for the SZ effect towards massive clusters. 

\vspace{0.5cm}

High resolution versions of the maps shown are available from the website: \\
http://www-astro.physics.ox.ac.uk/$\sim$gxl/sz\underline{ }images/


\section*{Acknowledgements}
\label{sec:Ack}
 
We would like to thank Marian Douspis for useful discussions. JMD is 
supported by a Marie Curie Fellowship of the European Community programme 
\emph{Improving the Human Research Potential and Socio-Economic knowledge} 
under contract number HPMF-CT-2000-00967. HM acknowledges financial support from PPARC. 

\bibliographystyle{mnras}

\bsp
\label{lastpage}
\end{document}